\documentclass[review]{elsarticle}

\usepackage{lineno,hyperref}
\usepackage{amsmath}

\journal{Journal of \LaTeX\ Templates}

\begin{document}

\begin{frontmatter}	

\title{Geometric mean extension for data sets with zeros}


\author[mymainaddress]{Roberto de la Cruz\corref{mycorrespondingauthor}}
\cortext[mycorrespondingauthor]{Corresponding author}
\ead{robertodelacruzmoreno@gmail.com}

\author[mymainaddress]{Jan-Ulrich Kreft}
\ead{j.kreft@bham.ac.uk}

\address[mymainaddress]{Institute of Microbiology and Infection, School of Biosciences, University of Birmingham, Birmingham, UK}

\begin{abstract}
There are numerous examples in different research fields where the use of the geometric mean is more appropriate than the arithmetic mean. However, the geometric mean has a serious limitation in comparison with the arithmetic mean. Means are used to summarize the information in a large set of values in a single number; yet, the geometric mean of a data set with at least one zero is always zero. As a result, the geometric mean does not capture any information about the non-zero values. The purpose of this short contribution is to review solutions proposed in the literature that enable the computation of the geometric mean of data sets containing zeros and to show that they do not fulfil the `recovery' or `monotonicity' conditions that we define. The standard geometric mean should be recovered from the modified geometric mean if the data set does not contain any zeros (recovery condition). Also, if the values of an ordered data set are greater one by one than the values of another data set then the modified geometric mean of the first data set must be greater than the modified geometric mean of the second data set (monotonicity condition). We then formulate a modified version of the geometric mean that can handle zeros while satisfying both desired conditions.
\end{abstract}

\begin{keyword}
Arithmetic mean \sep Geometric mean \sep  Log-normal distribution \sep Zero values
\end{keyword}

\end{frontmatter}


\section{Introduction}

Increasingly, research generates large amounts of information. Yet it is hard to work with large data sets directly, hence it is necessary to reduce their complexity whilst maintaining essential information. The most common way in which data sets are summarised is the use of the statistical quantities called means, e.g., the arithmetic mean or the geometric mean. Such means summarize all the information of the data set in a single value. In particular, the geometric mean is widely used in research fields such as biological sciences \cite{thomas_what_1990}, environmental sciences \cite{hirzel_modeling_2003} and economics \cite{curran_valuing_1994}. 

One of the limitations of the geometric mean is that it is not useful for data sets that contain one or more zero values. By definition, the geometric mean of a set of numbers with at least one zero is always zero, hence, it does not capture any information about the non-zero values. Thus, it is necessary to formulate a modified version of the geometric mean that is applicable to sets with zeros. The purpose of this contribution is to review previously proposed solutions and to develop one that fulfils the desirable conditions of recovery and monotonicity that we define in Section \ref{s:conditions}. 

\section{Standard geometric mean}\label{s:intro}
The geometric mean \cite{abramowitz_handbook_1972} of a sequence $X=\{x_{i}\}_{i=1}^{n}$, $x_{i}>0$ , is defined as:

\begin{equation}
	G(X)=\left(\prod_{i=1}^{n}x_{i} \right)^{1/n} \label{geometricmean}
\end{equation}

When n is large, it is computationally more feasible to use the following equivalent expression of (\ref{geometricmean}):

\begin{equation}
	G(X)=\exp \left({\frac{1}{n}\sum_{i=1}^{n} \log(x_{i})}\right)\label{geometricmean2}
\end{equation}

There are several reasons why using the geometric mean can be preferable to using the arithmetic mean. One important case is when the values in a set range over several orders of magnitude (e.g., plasmid transfer rates range over 7 orders of magnitude \cite{baker_mathematical_2016}), since the arithmetic mean would be dominated by the largest values. This makes application of the arithmetic mean to such data sets meaningless. In particular, values from a log-normal distribution have that feature. This distribution is important since the product of positive random independent variables (with  square-integrable density function) tends toward such a distribution as a consequence of the central limit theorem \cite{hazewinkel_encyclopaedia_1990}. While normal distributions arise from many small additive errors, log-normal distributions arise from multiplicative errors, which are common for growth processes or others with positive feedback \cite{limpert_log-normal_2001}. They are therefore common in biological, environmental and economical sciences, i.e., as growth of organisms, populations or assets is proportional to their current values, any extra increase by chance will multiply further increases. 

\section{Desired conditions of the geometric mean extension}\label{s:conditions}
There are two conditions that any extension of the geometric mean $G$ (that we denominate by $G_{0}$) should satisfy:   
\begin{itemize}
\item
Recovery condition. The usual geometric mean should be recovered when there are no zero-values, i.e., the relative difference between the standard geometric mean and its extension for a data set without zero values should be small.
If $X$ is a data set of only positive values,  $G_{0}(X) \simeq G(X)$.
\item
Monotonicity condition.
$G_{0}$ is monotone non-decreasing in the data set, i.e., if the values of a data set ($X_{1}$) are greater one by one than the values of another data set ($X_{2}$), then  $G_{0}(X_{1}) \geq G_{0}(X_{2})$.
In particular, the modified geometric mean should never increase when adding zeros to a data set.
\end{itemize}

\section{Habib's proposed solution}
\label{s:habib}
Only two solutions to this problem have been proposed to date and none satisfy both aforementioned conditions.\citep{habib_geometric_2012} proposes a geometric mean expression for probability distributions whose domains include zero and/or negative values and derives an expression to calculate the geometric mean of such a data set. In particular, if we have a data set $X=\{x_{1},x_{2},\cdots,x_{n},0,0,\cdots,0\}$ with $n$ positive values and $m$ zeros, 
then the geometric mean of $X$ according to \citep{habib_geometric_2012} is:

\begin{equation}
G_{1} (X)=\frac{n}{n+m} \exp{\left(\frac{1}{n+m}\sum_{i=1}^{n}\log(x_{i})\right)} \label{eqgeommeanHabib}
\end{equation}

\begin{figure}
    \centering
    \includegraphics[width=0.8\textwidth]{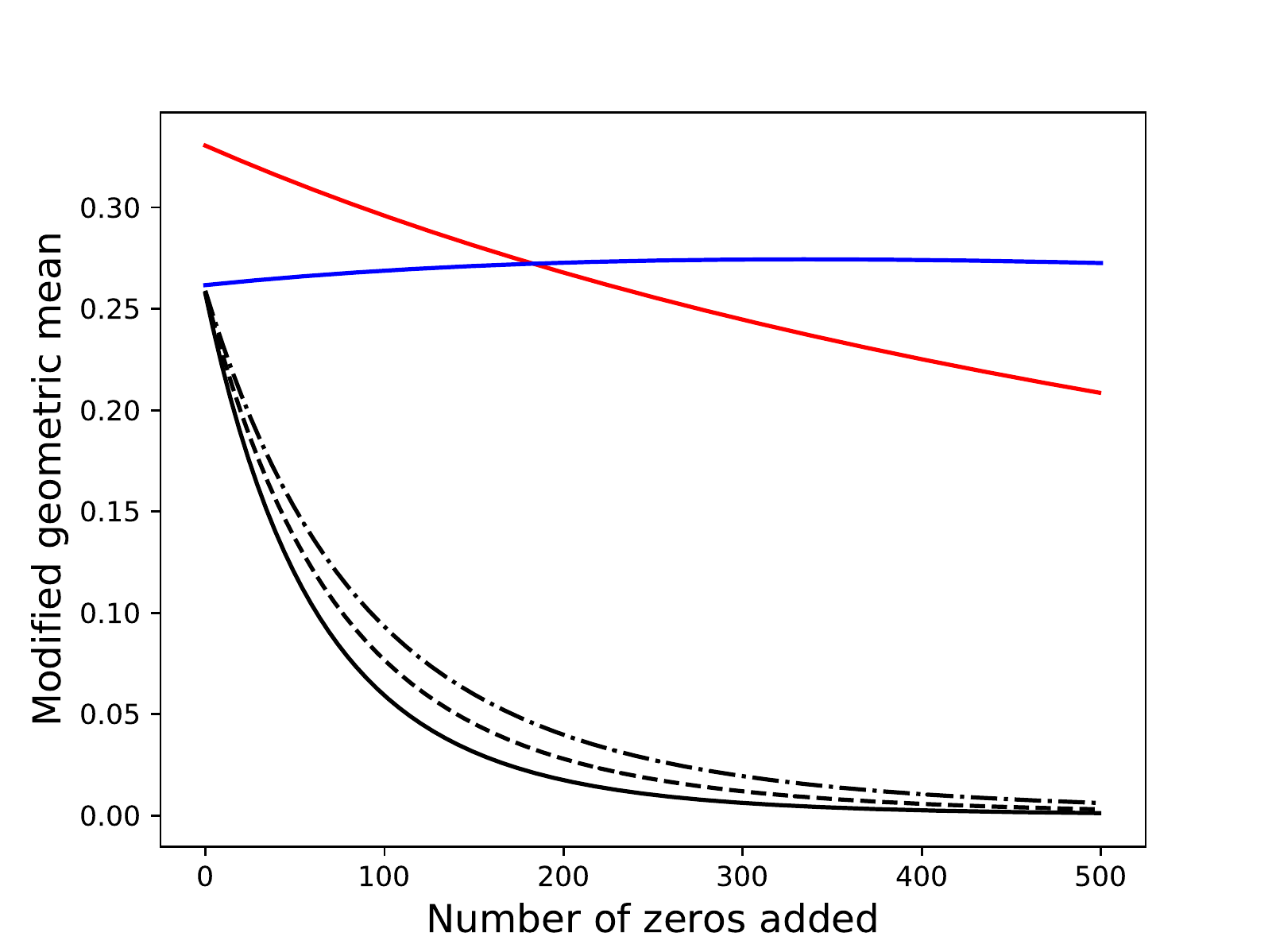}
    \caption{Different modified geometric means as a function of the number of zeros added to a data set of 983 positives values. These data come from simulations of Robyn Wright (University of Warwick) and represent specific growth rates of bacteria of different ages, including zero values. The original data set contains 1000 values (983 positive and 17 zero). The figure illustrates how Habib's modified geometric mean (blue line) is increasing despite adding zeros, before declining after 300 zeros have been added. The red line represents the solution described in Section $\ref{s:plus1}$, which does not recover the standard geometric mean (Recovery property). The three black lines represent our solution (Section $\ref{s:solution}$), for different values of $\epsilon$ (solid line $\epsilon=10^{-5}$, dashed line $\epsilon=10^{-4}$, dashdotted line $\epsilon=10^{-3}$) }
    \label{fig:habib}
\end{figure}

$G_{1}(X_{+}) < G_{1}(X)$. Therefore, it is not a desirable solution. We show an example of this with real data in Figure \ref{fig:habib}.

\section{An often used solution is to add one to all data in a set}\label{s:plus1}
Another proposed solution to deal with a data set with zeros is to add one to each value in the data set, calculating the geometric mean of this shifted data set according to (\ref{geometricmean2}) and then subtracting one again:

\begin{equation}
\exp{ \left(\frac{1}{n}\sum_{i=1}^{n} \log(x_{i}+1)\right)}-1, \label{geometricmeanadd1}
\end{equation}

This is a frequently used workaround in a number of different areas such as epidemiology \cite{alexander_index_2005,calhoun_combined_2007,barnard_measurement_2014,emerson_effect_1999,naish_prevalence_2004}, human pharmacology \cite{petry_efficacy_2013,el_setouhy_randomized_2004}, veterinary pharmacology \cite{shoop_discovery_2014,tielemans_comparative_2010}, entomology \cite{williams_time_1935,williams_use_1937}, marine ecology \cite{kuhn_plastic_2012}, environmental technology \cite{bastos_giardia_2004} and sociology \cite{thelwall_goodreads:_2017}).

The problem with this workaround (\ref{geometricmeanadd1}) is the arbitrariness of adding 1. As a consequence, (\ref{geometricmeanadd1}) does not satisfy the Recovery condition.

\section{Our solution and its problem and how to solve this}\label{s:solution}
Nevertheless, we can exploit the idea in section \ref{s:plus1} to formulate a modified version of the geometric mean, based on adding are optimal value $\delta$ to all data in a set that optimizes the difference between the standard geometric mean and the modified version for the subset of positive values.

Let $X=\{x_{1}, x_{2}, \cdots, x_{n}\}$, $x_{i} \geq 0 \ \forall i$ and $\exists  j \  x_{j}=0$; let $X_{+}=\{ x \in X \rvert x>0 \}$ . We define $G_{\epsilon,X}$ as

\begin{equation}
G_{\epsilon,X}(X)=\exp{ \left(\frac{1}{n}\sum_{i=1}^{n} \log(x_{i}+\delta)\right)}-\delta \label{modifiedgeometricmean}
\end{equation}

\noindent where 

 \begin{equation}
 \begin{array}{l}
 \delta=\sup \{ \delta_{*} \in (0,\infty) \ \rvert \  (G_{\epsilon,X}(X_{+})-G(X_{+}))¦< \epsilon G(X_{+}) \}, \label{deltacondition}\quad 0<\epsilon \ll 1
 \end{array}
 \end{equation} 

$\epsilon$ denotes the maximum relative difference between the standard geometric mean and our modified version for the subset of positive values. If not all values of $X$ are the same, $\delta$ is well-defined because  $G_{\epsilon,X}(X)$ is  strictly increasing in $\delta$ as a consequence of the superadditivity of the geometric mean. In the trivial case where all values of $X$ are equal to $x$, we could directly define $G_{\epsilon,X}(X) = x$. This modified geometric mean fulfils the Recovery and Monotonicity
requirements.

Note that $G_{\epsilon,X}$ depends on the data set $X$. This is a problem when comparing data sets. In order to deal with this inconvenience, we propose the following procedure:

Let $\epsilon>0$ and $X_{1}$, $X_{2}$, $\cdots$, $X_{n}$ the data sets to compare. For each $X_{i}$ we calculate $\delta_{i}$ according to (\ref{deltacondition}). Let $\delta_{min}=\min\{\delta_{1},\delta_{2},\cdots,\delta_{n}\}$. As $\delta_{min} \leq \delta_{i}$ $\forall i$, and (\ref{modifiedgeometricmean}) is increasing in $\delta$, $\delta_{min}$ satifies the condition expressed in (\ref{deltacondition}) for each data set. Using this $\delta_{min}$, we can calculate ($\ref{modifiedgeometricmean}$) for each $X_{i}$ in a unified way.

\section{Algorithm}
\label{s:alg}

\textit{Inputs:} $X$ (data set), $\epsilon$
\begin{tabbing}
 \qquad \enspace Calculate $X_{+} \subset X$, the set of the positive values of $X$ \\
 \qquad \enspace Calculate the geometric mean of $X_{+}$  \\
 \qquad \enspace By the bisection, calculate $\delta$ of ($\ref{deltacondition}$)\\
 \qquad \enspace Calculate $G_{\epsilon,X}(X)$ using $\delta$
 \end{tabbing}

\section{Geometric standard deviation}
\label{s:gsd}
If the geometric mean of a sequence, $X=\{x_{i}\}_{i=1}^{n}$, $x_{i}>0$, is denoted as $G$, then the geometric standard deviation \cite{kirkwood_geometric_1979} is defined as:

\begin{equation}
\sigma_{gsd}=\exp\left({\sqrt{\frac{\sum_{i=1}^{n}\left(\log (x_{i}/G) \right)^{2}}{n}}}\right) \label{eqgsd}
\end{equation}

Note that unlike the standard deviation, the geometric standard deviation is a multiplicative factor. That is, instead of subtracting and adding the arithmetic standard deviation to the arithmetic mean to obtain the lower and upper bounds of the interval, the geometric mean is divided and multiplied by the geometric standard deviation to obtain the upper and lower bounds of the interval.  As in the case of the geometric mean, the geometric standard deviation is not useful for data sets with zero values (Eq. (\ref{eqgsd}) is not even defined in this case). 

Nevertheless, we can, analogously to section \ref{s:solution}, formulate a modified version of (\ref{eqgsd}) in order to include data with zero values:
Let $X=\{x_{1}, x_{2}, \cdots, x_{n}\}$, $x_{i} \geq 0 \ \forall i$ and $\exists  j \  x_{j}=0$. We define $\sigma_{gsd}^{\epsilon,X}$ as:

\begin{equation}
\sigma_{gsd}^{\epsilon,X}=\exp\left({\sqrt{\frac{\sum_{i=1}^{n}\left(\log ((x_{i}+\delta_{*})/G_{\epsilon,X}) \right)^{2}}{n}}}\right) \label{eqgsdmod}
\end{equation}

where $\epsilon, \delta_{*}$ and $G_{\epsilon,X}$ have the same meaning as in section \ref{s:solution}.

\begin{figure}
    \centerline{\includegraphics[width=.8\textwidth]{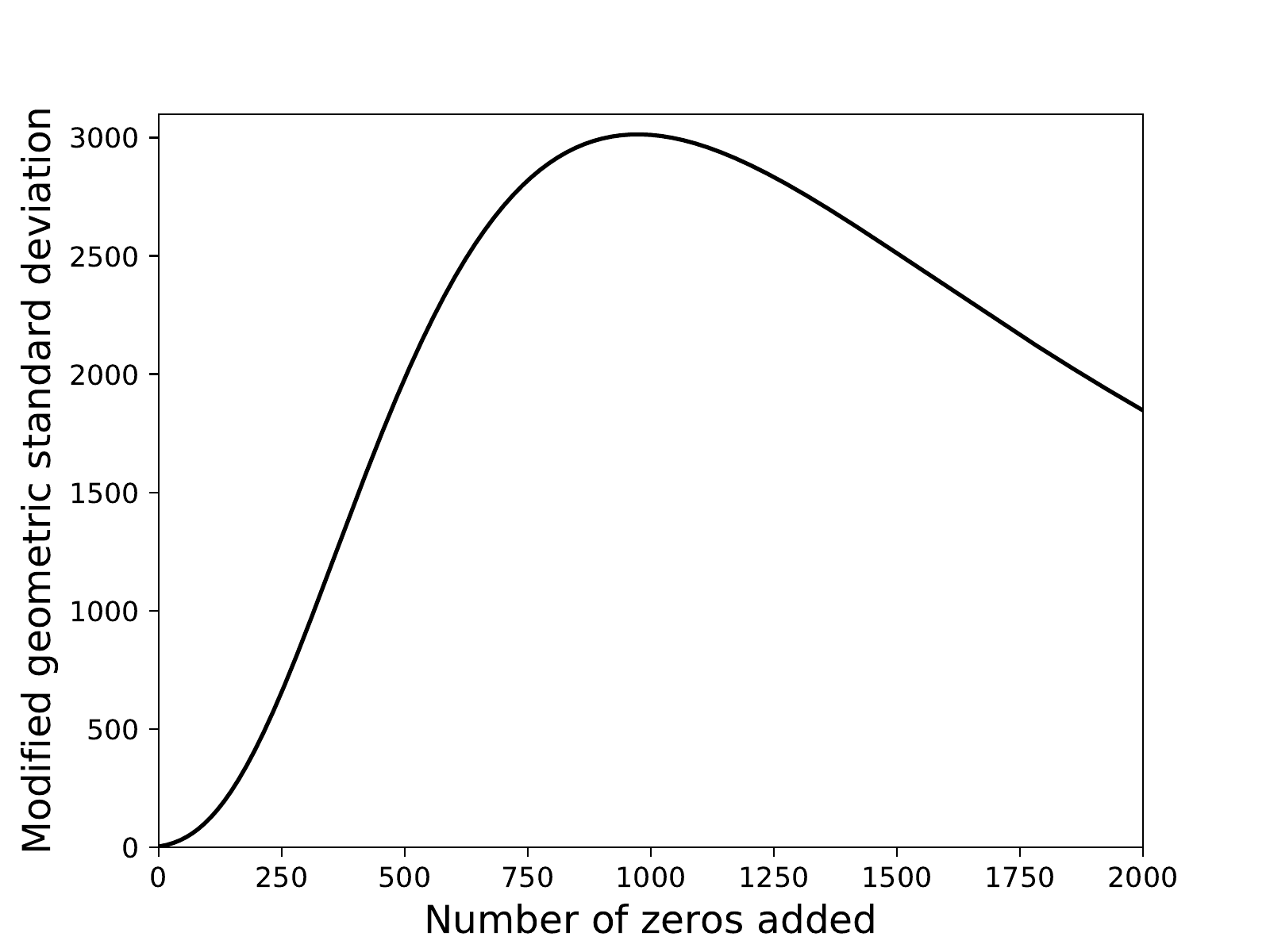}}
    \caption{Modified geometric standard deviation as a function of the number of zeros added to a data set of 983 positives values (same data as in Fig \ref{fig:habib}). The modified geometric standard deviation increases strongly when zeros are added from a value of $\approx$ 3 without zeros, and eventually declines again.}
    \label{fig:stdmod}
\end{figure}

As an example, we can use the same data as in Fig. \ref{fig:habib} to calculate how the modified geometric standard deviation varies depending on the number of zeros added (Fig. \ref{fig:stdmod}). Figure \ref{fig:stdmod} shows that our modified geometric standard deviation could take excessively large values, so our proposal is not ideal and we suggest that finding a better solution for the geometric standard deviation is an open problem.

\section{Discussion}
\label{s:discuss}
The geometric mean is preferable to the arithmetic mean when data are log-normally distributed or range over orders of magnitude; but the geometric mean has an important drawback when data sets contain zeros. Here, we propose a modified version of the geometric mean that can be used for data sets containing zeros. Previously proposed solutions have undesired properties that our solution avoids. 

However, our proposed solution is not universal (it depends on the data set). Likewise, the solution for calculating the geometric standard deviation that we propose is not ideal as it can lead to excessively large values. We hope that our little contribution stimulates the search for a universal solution. Nonetheless, we have developed a procedure to compare different data sets using the modified geometric mean $G_{\epsilon,X}$. Our open source Matlab and Python code is available on GitHub (\emph{https://github.com/RobertoCM/geomMeanExt}).

\section*{Acknowledgement}
The authors thank Robyn Wright (University of Warwick) for posing the problem and JPIAMR/MRC for funding (Dynamics of Antimicrobial Resistance in the Urban Water Cycle in Europe (DARWIN), MR/P028 195/1).


\section*{References}

\begin{thebibliography}{10}
\expandafter\ifx\csname url\endcsname\relax
  \def\url#1{\texttt{#1}}\fi
\expandafter\ifx\csname urlprefix\endcsname\relax\def\urlprefix{URL }\fi
\expandafter\ifx\csname href\endcsname\relax
  \def\href#1#2{#2} \def\path#1{#1}\fi

\bibitem{thomas_what_1990}
C.~D. Thomas, What do real population dynamics tell us about minimum viable
  population sizes?, Conservation Biology 4~(3)  324--327.

\bibitem{hirzel_modeling_2003}
A.~H. Hirzel, R.~Arlettaz, Modeling habitat suitability for complex species
  distributions by environmental-distance geometric mean, Environmental
  Management 32~(5)  614--623.

\bibitem{curran_valuing_1994}
M.~Curran, Valuing asian and portfolio options by conditioning on the geometric
  mean price, Management Science 40~(12)  1705--1711.

\bibitem{abramowitz_handbook_1972}
M.~Abramowitz, I.~A. Stegun, Handbook of Mathematical Functions: with Formulas,
  Graphs, and Mathematical Tables,, Dover Publications, Inc.

\bibitem{baker_mathematical_2016}
M.~Baker, J.~L. Hobman, C.~E.~R. Dodd, S.~J. Ramsden, D.~J. Stekel,
  Mathematical modelling of antimicrobial resistance in agricultural waste
  highlights importance of gene transfer rate, {FEMS} Microbiology Ecology
  92~(4)  fiw040.

\bibitem{hazewinkel_encyclopaedia_1990}
Central limit theorem, Encyclopaedia of Mathematics, Springer Netherlands.

\bibitem{limpert_log-normal_2001}
E.~Limpert, W.~A. Stahel, M.~Abbt, Log-normal distributions across the
  sciences: Keys and clues, {BioScience} 51~(5)  341--352.

\bibitem{habib_geometric_2012}
E.~A.~E. Habib, Geometric mean for negative and zero values, International
  Journal of Research and Reviews in Applied Sciences 11~(3)  419--32.

\bibitem{alexander_index_2005}
N.~Alexander, A.~Solomon, M.~Holland, R.~Bailey, S.~West, J.~Shao, D.~Mabey,
  A.~Foster, An index of community ocular chlamydia trachomatis load for
  control of trachoma, Transactions of the Royal Society of Tropical Medicine
  and Hygiene 99~(3)  175--177.

\bibitem{calhoun_combined_2007}
L.~M. Calhoun, M.~Avery, L.~Jones, K.~Gunarto, R.~King, J.~Roberts, T.~R.
  Burkot, Combined sewage overflows ({CSO}) are major urban breeding sites for
  culex quinquefasciatus in atlanta, georgia, The American Journal of Tropical
  Medicine and Hygiene 77~(3)  478--484.

\bibitem{barnard_measurement_2014}
D.~R. Barnard, C.~Z. Dickerson, K.~Murugan, R.-D. Xue, D.~L. Kline, U.~R.
  Bernier, Measurement of landing mosquito density on humans, Acta Tropica 136
  58--67.

\bibitem{emerson_effect_1999}
P.~M. Emerson, S.~W. Lindsay, G.~E. Walraven, H.~Faal, C.~Bøgh, K.~Lowe, R.~L.
  Bailey, Effect of fly control on trachoma and diarrhoea, The Lancet
  353~(9162)  1401--1403.

\bibitem{naish_prevalence_2004}
S.~Naish, J.~{McCarthy}, G.~Williams, Prevalence, intensity and risk factors
  for soil-transmitted helminth infection in a south indian fishing village,
  Acta Tropica 91~(2)  177--187.

\bibitem{petry_efficacy_2013}
G.~Petry, G.~Altreuther, S.~Wolken, P.~Swart, D.~J. Kok, Efficacy of emodepside
  plus toltrazuril oral suspension for dogs
  (procox\textsuperscript{\textregistered}, bayer) against trichuris vulpis in
  naturally infected dogs, Parasitology Research 112  133--138.

\bibitem{shoop_discovery_2014}
W.~L. Shoop, E.~J. Hartline, B.~R. Gould, M.~E. Waddell, R.~G. {McDowell},
  J.~B. Kinney, G.~P. Lahm, J.~K. Long, M.~Xu, T.~Wagerle, G.~S. Jones, R.~F.
  Dietrich, D.~Cordova, M.~E. Schroeder, D.~F. Rhoades, E.~A. Benner, P.~N.
  Confalone, Discovery and mode of action of afoxolaner, a new isoxazoline
  parasiticide for dogs, Veterinary Parasitology 201~(3)  179--189.

\bibitem{tielemans_comparative_2010}
E.~Tielemans, C.~Manavella, M.~Polmeier, T.~Chester, M.~Murphy, B.~Gale,
  Comparative acaricidal efficacy of the topically applied combinations
  fipronil/(s)-methoprene, permethrin/imidacloprid and metaflumizone/ amitraz
  against \textit{Dermacentor reticulatus,} the european dog tick (ornate dog
  tick, fabricius, 1794) in dogs, Parasite 17~(4)  343--348.

\bibitem{kuhn_plastic_2012}
S.~K\~uhn, J.~A. van Franeker, Plastic ingestion by the northern fulmar
  (fulmarus glacialis) in iceland, Marine Pollution Bulletin 64~(6)
  1252--1254.

\bibitem{bastos_giardia_2004}
R.~Bastos, L.~Heller, M.~Vieira, L.~Brito, P.~Bevilacqua, L.~Nascimento,
  Giardia sp. cysts and cryptosporidium spp. oocysts dynamics in southeast
  brazil: occurrence in surface water and removal in water treatment processes,
  Water Science and Technology: Water Supply 4~(2)  15--22.

\bibitem{thelwall_goodreads:_2017}
M.~Thelwall, K.~Kousha, Goodreads: A social network site for book readers,
  Journal of the Association for Information Science and Technology 68~(4)
  972--983.

\end{thebibliography}
\end{document}